\documentclass[twocolumn,twoside,slac_two]{revtex4}
\usepackage{graphicx}
\usepackage{fancyhdr}
\begin{document}

\title{Diboson Production at D\O{}}

%

\author{M.~A. Strang\footnote
{\uppercase{O}n behalf of the \uppercase{D}\O{} \uppercase{C}ollaboration.}}
\affiliation{Department of Physics, The Ohio State Universtity, Columbus, \\
OH 43210, USA}

\begin{abstract}
We present high statistics measurements of diboson production in ppbar 
collisions at the D\O{} experiment in multiple channels, including 
$ZZ \rightarrow llll$, 
$Z\gamma \rightarrow \nu\nu\gamma$, 
$WW \rightarrow l \nu l \nu$ 
and $WW/WZ \rightarrow l \nu jj$. 
These measurements both test physics beyond the 
standard model and demonstrate the sensitivity of hadron colliders to rare 
signals such as Higgs boson production.
\end{abstract}

\maketitle

\thispagestyle{fancy}


\section{Introduction}
The study of diboson physics at the Tevatron allows us to probe gauge
boson self-interactions that are a consequence of the non-Abelian nature of
the SU(2)$_{L}$ $\otimes$ U(1)$_{Y}$ symmetry group. This is one of the least 
tested areas of the Standard Model.

Studies in this area provide a natural series of goals for detector 
sensitivity, probe fundamental details of the standard model electroweak
sector directory and measurements at the Tevatron explore higher energies
than LEP while providing access to some channels ({\it e.g.} $WZ$) not
available there. New physics would be reflected in increased cross sections
beyond standard model expectations.

Diboson physics also form backgrounds to many other interesting physics
channels like Higgs, SUSY and ttbar, so understanding of the diboson
physics can help studies at the LHC and Tevatron.

In this proceedings, we present the results from four analyses.

\section{D\O{} Experiment}
D\O{} is a multipurpose, hermetic detector. The interaction region is 
surrounded by a silicon microstrip detector and a fiber tracker within a
2~T superconducting solenoid magnet. There is a liquid argon - uranium 
calorimeter behind the magnet to measure the energy of most particles. 
The calorimeter is surrounded by three layers of muon detector, with a 1.8~T
toroidal magnet after the first layer. It has pseudo-rapidity coverage of up 
to 3.2 for electrons and up to 2.0 for muons. The detector is completely 
described in the corresponding D\O{} NIM article.~\cite{d0detector}

\section{$ZZ \rightarrow llll$ Production}
The SM prediction for the total $ZZ$ production cross section in $p\bar{p}$
collisions at the Fermilab Tevatron Collider at $\sqrt{s}=1.96$~TeV is
$\sigma(ZZ)=1.4 \pm 0.1$~pb~\cite{Campbell:1999ah}. The requirement of 
leptonic $Z$ boson decays -- each $Z$ boson decay to either electron or muon 
pairs, resulting in final states consisting of four electrons ($4e$), four 
muons ($4\mu$) or two muons and two electrons ($2\mu2e$) -- reduces the 
observable cross section, making its 
measurement rather challenging. The accumulation of integrated luminosities 
in excess of 3 fb$^{-1}$ at the Fermilab Tevatron Collider and the development 
of highly optimized event selection criteria has now made possible the direct 
observation of $ZZ$ production using a data set of 1.7 fb$^{-1}$. 

For the $4e$ channel, we require the measurement of four electrons with ordered
transverse energies $E_{T} >$ 30, 25, 15, and 15~GeV, respectively, sorted
by the number of electrons in the central detector in order to exploit 
differences in the QCD background, requiring at least two electrons in 
the central detector. For the $4\mu$ channel, each measured muon 
must satisfy quality criteria based on scintillator and wire chamber 
information from the muon system, and have a matched track in the central 
tracker. We require that the four most energetic muons have ordered
transverse momenta $p_T >$ 30, 25, 15, and 15~GeV, respectively.
At least three muons in the event must be isolated. For the $2\mu2e$ channel, 
the two most energetic electrons and muons in an event must have 
$E_T(p_T)> 25, 15$ GeV sorted by the number of electrons in the central 
detector. In all cases, the leptons 
have to be consistent with coming from a pair of $Z$ bosons with one having
a dilepton mass greater than 70 GeV and the other greater than 50 GeV, with
opposite-charge, like flavor lepton pairing.

This selection provides for a clean signature with no other standard model
background with four leptons. The small number of expected events means
understanding of the background is important, the majority of which comes
from $Z/\gamma$ + jets, where the jets are misreconstructed as leptons. This
background is sensitive to the number of electrons in the central calorimeter.

\begin{figure}[ht]
\centering
\includegraphics[width=80mm]{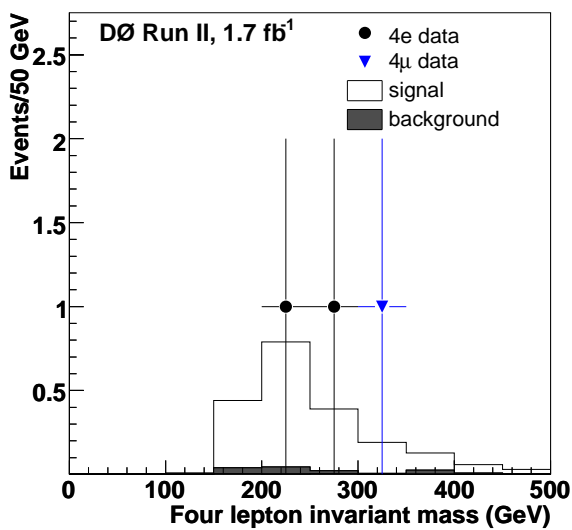}
\caption{Distribution of four lepton invariant mass in data,
expected signal, and expected background.} \label{zzfig}
\end{figure}

The predicted background is $0.14^{+0.03}_{-0.02}$ and the predicted signal
is $1.89\pm0.08$ with 3 events observed in the data as shown in 
Fig.~\ref{zzfig}. The likelihood of the
background fluctuating to give the observed yield ($p$-value) is $4.3 \times
10^{-8}$ corresponding to a significance of $5.3\sigma$. When this result
is combined with an earlier $ZZ$ analysis in the same channel and a
$ZZ \rightarrow ll\nu\nu$, we get a significance of $5.7\sigma$ and measure
a cross section of 
$1.60 \pm 0.63~\mathrm{(stat.)}^{+0.16}_{-0.17}~\mathrm{(syst.)}$~pb.~\cite{ZZ}
This is in agreement with the theoretical expectation.

\section{$Z\gamma \rightarrow \nu\nu\gamma$ Production}
The SM prediction for the $Z\gamma \rightarrow \nu\nu\gamma$ production cross 
section in $p\bar{p}$ collisions with a tight transverse energy cut on the 
photon of 90 GeV is 
$\sigma(Z\gamma) /dot \mathrm{BR}(Z/rightarrow /nu/nu) = 39 \pm 4$~pb
\cite{Zgammatheory}. The measurement in this channel is performed with a
data set of 3.6 fb$^{-1}$. 

As stated, a single high $E_{T}$ photon is required in the central calorimeter
and a large missing transverse energy of greater than 70 GeV is also required
to suppress multijet background. Backgrounds are further reduced by
rejecting events with jets having $p_{T} > 15$ GeV which contribute to
mismeasured missing $E_{T}$ and the background from $W \rightarrow l\nu$
and $Z \rightarrow ll$ is reduced by vetoing on muons, isolated tracks and
additional electromagnetic objects with a transverse momentum greater than
15 GeV. Non-collision backround ({\it e.g.} muon from halo or cosmics
undergoing bremsstrahlung) is handled using a pointing algorithm making use
of the energy deposited in the calorimeter as well as measured tracks, where
we assume the electromagnetic shower is initiated by photons, and require
that the vertex pointed to by the energy is within 10 cm of the actual 
vertex, and that the distance of closest approach for the track is within
4 cm of the actual vertex.

The predicted background is $17.3 \pm 2.4$ and the predicted signal
is $33.7\pm3.4$ with 51 events observed in the data. The likelihood of the
background fluctuating to give the observed yield ($p$-value) is $3.1 \times
10^{-7}$ corresponding to a significance of $5.1\sigma$. We measure
a cross section of 
$32.9 \pm 9~\mathrm{(stat.+syst.)}\pm2{(lumi.)}$~fb.~\cite{Zgamma}
This is in agreement with the theoretical expectation.

\section{$WW \rightarrow l \nu l \nu$ Production}
The SM prediction for the $WW \rightarrow l \nu l \nu$ production cross 
section in $p\bar{p}$ collisions is $12.0 \pm 7$~pb
\cite{Campbell:1999ah}. The measurement in this channel is performed with a
data set of 1 fb$^{-1}$. 

Signal selection requires two leptons from the same vertex consisting of $ee$,
$e\mu$ or $\mu\mu$ of opposite charge with at least one electron in the
central calorimeter if appropriate and the leading lepton is required to have 
a transverse
momentum of 25 GeV while the other has 15 GeV.
Backgrounds are reduced using various additional cuts. The background from
$Z \rightarrow ll$ is reduced via optimized missing $E_{T}$ for each
channel, where the value is required to be greater than 44 GeV for $ee$,
20 GeV for $e\mu$ or 35 GeV for $\mu\mu$. This is further refined by an
invarient mass requirement in the $ee$ channel and azimuthal angle requirements
in the other two channels. To reduce top and $W$+jets backgrounds,
we require a balanceed event where $q_{T}$ (the vector sum of the transverse
momenta with the transverse missing energy) is reqired to be less than
20 GeV for $ee$, 25 GeV for $e\mu$ or 16 GeV for $\mu\mu$.

\begin{figure}[ht]
\centering
\includegraphics[width=40mm]{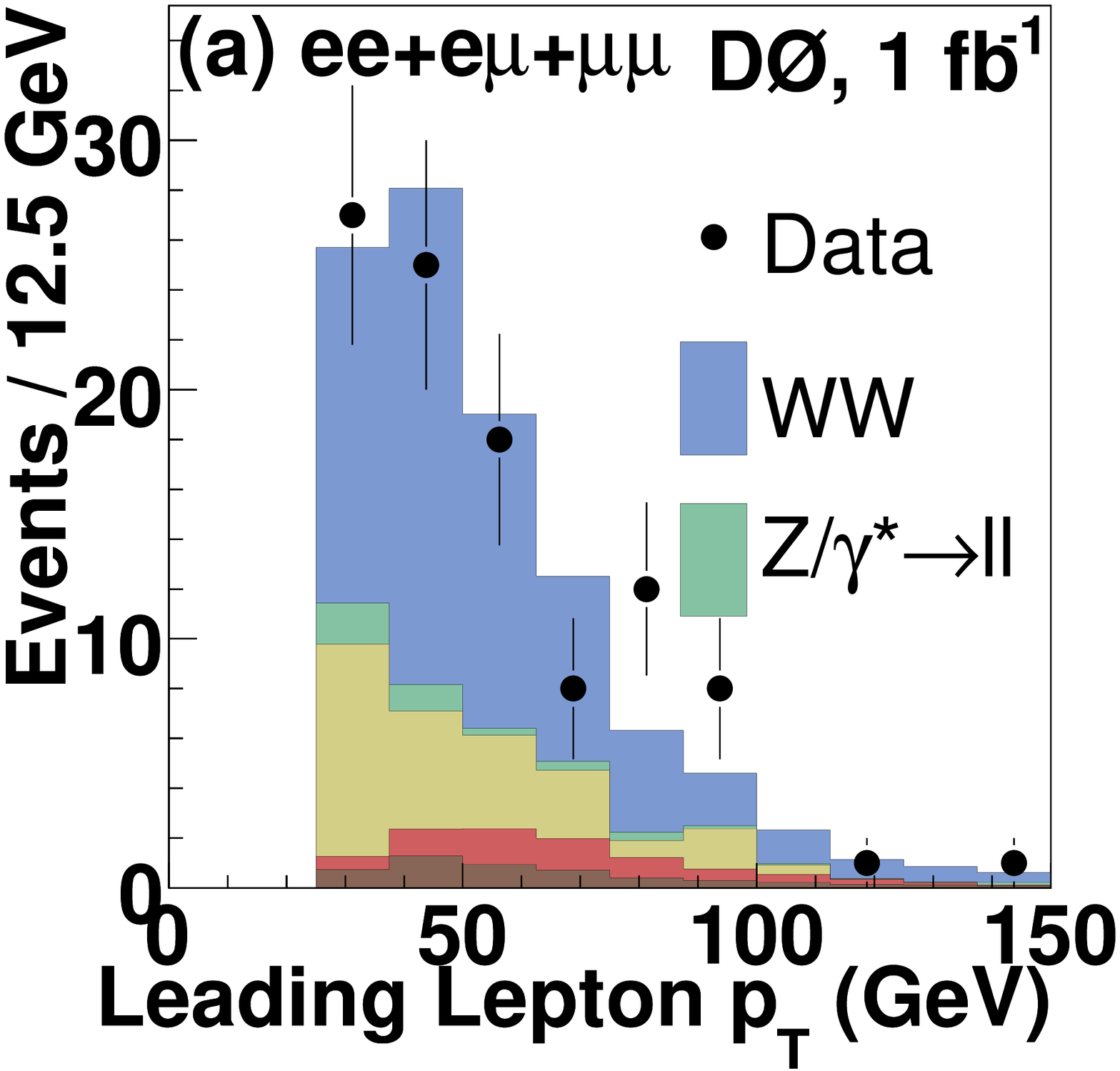}
\includegraphics[width=40mm]{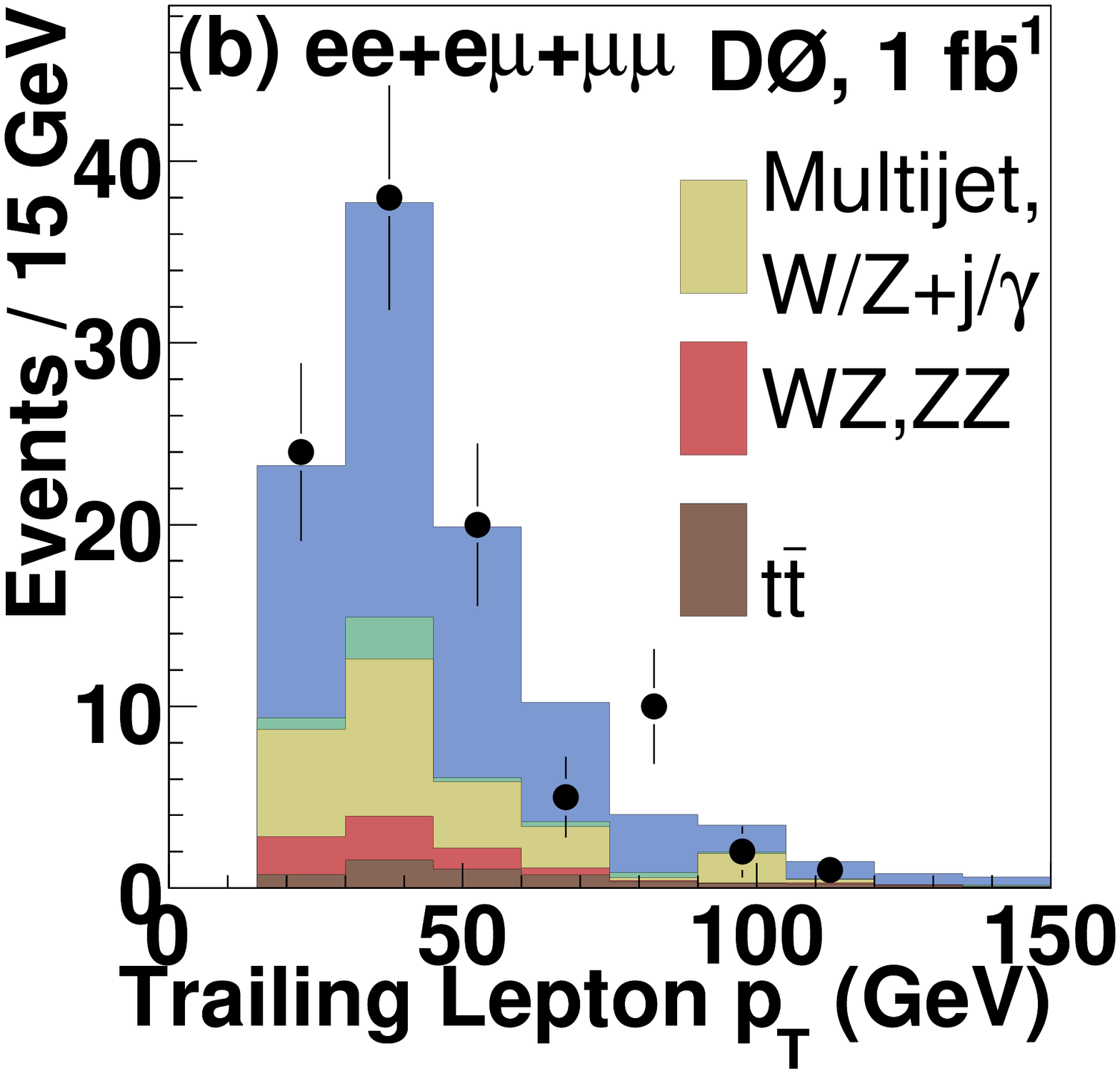}
\caption{Distributions of the (a) leading and (b) trailing
lepton $p_{T}$ after final selection, combined for all channels
($ee + e\mu + \mu\mu$). Data are compared to estimated signal,
$\sigma(WW) = 12$ pb, and background sum.} \label{wwfig}
\end{figure}

We measure a cross section of 
$11.5 \pm 2.1~\mathrm{(stat.+syst.)}\pm0.7{(lumi.)}$~pb.~\cite{WW}
This is in agreement with the theoretical expectation. Distributions in $p_{T}$
for the leading and trailing lepton are shown in Fig.~\ref{wwfig}.

\section{$WW/WZ \rightarrow l \nu jj$ Production}
The SM prediction for the $WW/WZ \rightarrow l \nu jj$ production cross 
section in $p\bar{p}$ collisions is $16.1 \pm 0.9$~pb
\cite{Campbell:1999ah}. The measurement in this channel is performed with a
data set of 1.1 fb$^{-1}$. 

\begin{figure}[ht]
\centering
\includegraphics[width=80mm]{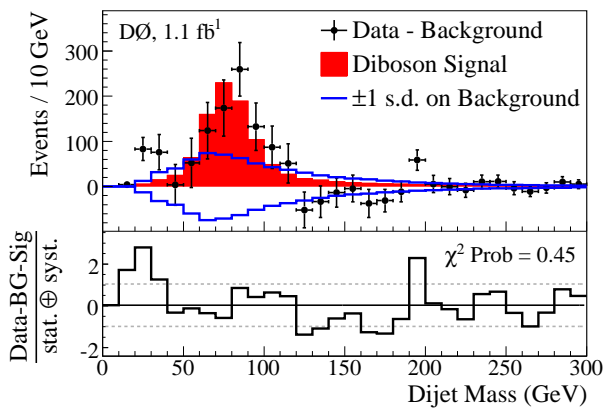}
\caption{A comparison of the extracted signal (fillled histogram)
to background-subtracted data (points), along with
the $\pm 1$ standard deviation (s.d.) systematic uncertainty on
the background. The residual distance between the data
points and the extracted signal, divided by the total uncertainty,
is given at the bottom.} \label{ww_wzfig}
\end{figure}

Signal selection requires one isolated lepton with $p_{T} > 20$ GeV in the
central calorimter, missing transverse energy greater than 20 GeV and two
jets, the first with $p_{T} > 30$ GeV and the second greater than 20 GeV.
Multijet backgrounds are reduced by requring a transverse $W$ mass greater
than 35 GeV, and backgrounds from $W$+jets, $Z$+jets, top etc. are reduced
using a ``Random Forest'' multivariate discriminant. The dijet mass peak 
extracted from data compared to the $WW=WZ$ MC prediction is
shown in Fig.~\ref{ww_wzfig}.

The likelihood of the background fluctuating to give the 
observed yield is $5.4 \times
10^{-6}$ corresponding to a significance of $4.4\sigma$. We measure
a cross section of 
$20.2 \pm 4.4~\mathrm{(stat.+syst.)}\pm1.2{(lumi.)}$~pb.~\cite{WZ}
This is in agreement with the theoretical expectation.

\section{Summary}
We have shown measurements for four different diboson processes. This includes
the first observation of $Z\gamma \rightarrow \nu\nu\gamma$ production, 
evidence for $WW/WZ \rightarrow l \nu jj$ production, a new measurement of 
$WW \rightarrow l \nu l \nu$ production and observation of $ZZ$ production.
So far, everything agrees with standard model expectations. With over 6 
fb$^{-1}$ now reconstructed, there will be more refined measurements in the
future.

\bigskip
\begin{acknowledgments}
We thank the staffs at Fermilab and collaborating institutions, 
and acknowledge support from the 
DOE and NSF (USA);
CEA and CNRS/IN2P3 (France);
FASI, Rosatom and RFBR (Russia);
CNPq, FAPERJ, FAPESP and FUNDUNESP (Brazil);
DAE and DST (India);
Colciencias (Colombia);
CONACyT (Mexico);
KRF and KOSEF (Korea);
CONICET and UBACyT (Argentina);
FOM (The Netherlands);
STFC and the Royal Society (United Kingdom);
MSMT and GACR (Czech Republic);
CRC Program, CFI, NSERC and WestGrid Project (Canada);
BMBF and DFG (Germany);
SFI (Ireland);
The Swedish Research Council (Sweden);
and
CAS and CNSF (China).
\end{acknowledgments}

\bigskip 

\end{document}